\documentclass[aps,preprint,amssymb,12pt,floatfix]{revtex4}
\usepackage{subfigure}
\usepackage{graphicx}
\usepackage{rotating}
\usepackage{amsmath,amsthm}

\begin{document}
\title{Internal strain regulates the nucleotide binding site of the kinesin leading head}
\author{Changbong Hyeon and Jos{\'e} N. Onuchic}
\thanks{To whom correspondence should be addressed at: Center for Theoretical Biological Physics, 
University of California at San Diego, 9500 Gilman Drive, La Jolla, CA 92093-0374. E-mail: jonuchic@ucsd.edu}
\affiliation{Center for Theoretical Biological Physics\\ 
University of California at San Diego, La Jolla, California 92093-0374\\
\[\]
}
\date{\today}
\baselineskip = 22pt

\begin{abstract}
In the presence of ATP, kinesin proceeds along the protofilament of microtubule by alternated binding of two 
motor domains on the tubulin binding sites. 
Since the processivity of kinesin is much higher than other motor proteins, 
it has been speculated that there exists a mechanism for 
allosteric regulation between the two monomers. 
Recent experiments suggest that ATP binding to the leading head domain in kinesin is regulated by the rearward strain built on the neck-linker. 
We test this hypothesis by explicitly modeling a $C_{\alpha}$-based kinesin structure whose 
both motor domains are bound on the tubulin binding sites. 
The equilibrium structures of kinesin on the microtubule show disordered and ordered neck-linker configurations for the leading and the trailing head, respectively. 
The comparison of the structures between the two heads shows that 
several native contacts present at the nucleotide binding site in the leading head are less intact than those in the binding site of the rear head. 
The network of native contacts obtained from this comparison 
provides the internal tension propagation pathway, 
which leads to the disruption of the nucleotide binding site in the leading head. 
Also, using an argument based on polymer theory, we estimate the internal tension built on the neck-linker to be $f\approx (12-15)$ pN. 
Both of these conclusions support the experimental hypothesis.  
\end{abstract}
\maketitle

\section{Introduction}
Extensive interest has recently been devoted to the understanding of molecular motors, which play 
pivotal roles in cellular processes by performing mechanical work using 
the energy-driven conformational changes.
Kinesin, myosin, F$_1$-ATPase, GroEL, RNA polymerase, and ribosome 
belong to a group of biological machines that undergoes a series of conformational changes during the mechanochemical cycle 
where the molecular conformation is directly coupled to the chemical state of the ligand. 
Although substantial progress has been achieved in understanding the underlying physical principles that govern molecular motors during the last decade, 
major issues still remain to be resolved. 
Specifically, some of the outstanding questions are as follows: 
(a) How is the chemical energy converted into mechanical work?
(b) How is the directionality of the molecular movement determined?
(c) How is the molecular movement coordinated or regulated?
Several biochemical experiments have quantified the kinetic steps \cite{Ma97JBC,Moyer98BC}, 
single molecule experiments using optical tweezers 
have measured the mechanical response of individual molecular motors \cite{Visscher99Science,BlockNature03,Chemla05Cell}, 
and an increasing number of crystal structures  have 
provided glimpses into the mechanisms of  molecular motors \cite{Mandelkow97Cell,HirokawaSCI04,Sablin98Nature,Rayment93SCI,Abrahams94Nature,SiglerNature97,Kornberg01Science,NollerSCI01}. 
These experimental evidences, however, are not sufficient to 
fully address all the questions above. For example, 
little is known not only about the structural details of  each chemical state but also about the kinetic pathways connecting them. 
Hence, if feasible, a computational strategy using the coordinates from X-ray and/or NMR structures 
can shed light on the allosteric dynamics of molecular motors. 
Although some initial numerical studies \cite{Koga06PNAS,Okazaki06PNAS,Hyeon06PNAS,Yu06BJ} have proceeded towards addressing issue (a) for a few cases where both open and closed structures are \emph{explicitly known}, no previous attempt has been made to answer issue (c). 
In this paper we investigate this question in the context of the 
conventional kinesin where the mechanochemical coordination of the motor movement is best manifested among the motor proteins.

One of the experimentally best studied molecular motors 
is the  conventional kinesin (kinesin-1) \cite{Brady85Nature,Vale85Cell}, a relatively small sized motor protein that transports cellular materials by walking in an unidirectional hand-over-hand manner along the microtubule (MT) filaments. 
Compared to other motor proteins involved in material transport such as myosin and dynein, 
the conventional kinesin has a remarkable processivity, which can travel about a hundred ($\sim$8.2 $nm$) steps without being dissociated from the MT. 
The mechanochemical cycle conjectured from experiments suggests that there must be a dynamic coordination between the two motor domains in order to achieve such high processivity. 
The quest to identify the origin of this dynamic coordination has drawn extensive attention among the kinesin community. 
Since Hancock and Howard \cite{Hancock99PNAS} first hypothesized that the ``internal strain'' was needed for processivity, 
the strain-dependent mechanochemistry became a popular subject in kinesin studies \cite{Uemura03NSB,GilbertPNAS04,BlockJBC03}.
With the aid of optical tweezers, Guydosh and Block recently revisited this issue by monitoring the real-time kinesin dynamics in the presence of ATP and $\mathrm{ADP\cdot BeF_x}$, 
a tight binding ATP analog \cite{BlockPNAS06}. 
They discovered that, when $\mathrm{ADP\cdot BeF_x}$ was bound to the kinesin, the pause-time of the step increased substantially and that the normal step was restored only after the 
obligatory backstep. This suggests that $\mathrm{ADP\cdot BeF_x}$ is 
released only when the head bound with $\mathrm{ADP\cdot BeF_x}$ becomes the leading head (L). 
Supported by this observation, they advocated a kinetic model in 
which the rearward strain via the neck-linker 
facilitates the release of the ligand from the L \cite{BlockPNAS06}.
Stated differently, the binding of the ligand to the L is inhibited because the 
rearward strain constitutes an unfavorable environment for the ATP binding sites of the L. In the present study,
we focus on the elucidation of the
structural origin of the coordinated motion in kinesin by adopting a simple 
computational strategy. 

Better straightforward evidence of the regulation on the nucleotide binding site 
can be obtained when a structure in which both kinesin heads are simultaneously bound to the MT binding site is determined. 
Such a structure will allow us to identify 
the structural differences between the leading (L) and the trailing (T) head. 
To date, this structure, however,  has not yet been reported. 
The only available structures include an isolated kinesin-1 without the MT \cite{Mandelkow97Cell},
an isolated single-headed kinesin-like (KIF1A) with various ligand states 
\cite{HirokawaSCI04}, and a single KIF1A bound to the tubulin-dimer binding site \cite{Kikkawa00Cell}. 
Therefore, we utilized existing Protein Data Bank (PDB) structures 
and manually built a model system of the  two-headed kinesin molecule 
with both heads bound to the tubulin binding sites (see Fig.2 and legend).
This model was used to generate an ensemble of structures via simulations. 
A direct comparison between the L and T equilibrium structures 
shows that the tension built on the neck-linker induces the disruption of the nucleotide binding site of the L, which directly supports inferences from experimental observations \cite{BlockJBC03,Uemura03NSB,GilbertPNAS04,BlockPNAS06}. \\

\section{Results and Discussion}

{\bf Mechanochemical cycle of kinesin: }
We begin by reviewing the mechanochemical cycle of kinesin molecule on the MT to clarify the importance of dynamic coordination between the two motor domains for kinesin processivity. 
Recent experiments using laser optical tweezers (LOT), cryo-electron microscopy, electron paramagnetic resonance, and FRET, as well as the crystal structures at various states \cite{HirokawaSCI04,ValeNature99,Sindelar02NSB,Cross04TBS} provide glimpses into the structural and dynamical details of how the kinesin molecule walks on the microbubule filaments.  
Depending on the nucleotide state at the binding site, 
both the motor domain structure and the binding interface between kinesin and MT are affected.
In particular, a minor change of the motor domain coupled to the nucleotide 
is amplified to a substantial conformational change 
of the neck linker between the ordered and the disordered state.  
Experimental studies strongly suggest the mechanochemical cycle 
shown in Fig.1 \cite{Cross04TBS}. 
The mechanical stepping cycle of kinesin initiates with the binding of ATP 
to the empty kinesin head strongly bound to the MT [$(i)\rightleftharpoons(ii)$]. 
Docking of the neck linker to the neck-linker binding motif on the leading head (X in Fig.1) propels the trailing 
head (Y in Fig.1) in the (+)-direction of MT, which leads to an 8 $nm$ mechanical step [$(ii)\rightarrow (ii')$]. 
The interaction with the MT facilitates the dissociation of ADP from the catalytic core of L [$(ii')\rightarrow (iii)$]. 
ATP is hydrolyzed and produces $\mathrm{ADP\cdot P_i}$ state for the T [$(iii)\rightarrow (iv)$]. When $P_i$ is released and the trailing head is unbound from the MT, the half-cycle is completed [$(iv)\rightarrow (i)$]. 
The mechanical step is achieved in a hand-over-hand fashion 
by alternating the binding of the two motor domains (X and Y in Fig.1) to the MT \cite{Hackeny94PNAS,Block03Science}. 
High processivity of the kinesin requires this kinetic cycle to be stable (remain within the yellow box in Fig.1). 
A premature binding of the ATP to the leading head in the state of $(E:MT)$ 
should be prevented, i.e., the condition 
$k_{bi}^{(iii)}[ATP]/(k_r^{(iii)}+k_{diss}^{(iii)})$ and $k_{bi}^{(iv)}[ATP]/(k_r^{(iv)}+k_{diss}^{(iv)})\rightarrow 0$ 
should be satisfied in Fig.1 (see Supporting Information for the master equation describing the kinetic cycle). 
ATP binding to the (iii) or (iv) states can destroy 
the mechanochemical cycle of the kinesin. 
The binding of ATP on the leading head should be suppressed before the 
$\gamma$-$P_i$ is released 
from the T.
Otherwise, both heads become ADP-bound states, which have a weak binding affinity to the MT, and that leads to dissociation from the MT. 
Since the kinesin has a high processivity compared to other molecular motors, 
effective communication is required between the two heads regarding 
the chemical state of each of the partner motor domains. 
\\

{\bf Two-headed kinesin bound to the microtubule: }
In the absence of interactions with the MT, 
the individual kinesin monomers fold into identical conformations. 
To achieve its biological function, however, 
folding into the native structure alone is not sufficient. 
Coupled with the nucleotide and the MT, 
the two kinesin monomers in the dimeric complex need to alternate the acquisition of the native structure in a time-coordinated fashion for the uni-directional movement. 
The currently available three-dimensional structure (PDB id: 3kin, structure 2 in Fig.2-A), in which each monomer is in its native state, does not provide such a dynamic picture by failing in fulfilling the geometrical requirement of simultaneous bindings of both motor domain to the adjacent tubulin 
binding sites that have an 8-nm gap. 
The inspection of 3-D structure suggests that a substantial increase of the distance between the two motor domains can be gained by breaking a few contacts associated with the neck-linker ($\beta 9$, $\beta 10$) and the neck-linker binding site on the motor domain ($\beta 7$). 
To this end, we manipulated the 3-D structure of 3kin around the neck-linker of the L and created a temporary structure whose two heads bind to the MT binding sites simultaneously. 
Both L and T have energetic biases towards the identical native 
fold but the interactions with the tubulin binding sites adapt the dimeric 
kinesin structure into a different minimum structure, 
which is not known \emph{a priori}. 
We performed simulations (see Methods) to relax this initial structure 
and to establish the thermal equilibrium ensemble of the kinesin molecule 
on MT (see Fig.3-A).  
Transient dimeric kinesin conformations corresponding to the steps (iii) and (iv) during the cycle (Fig.1) allow us to investigate the structural deviation between L and T of kinesin molecule. 
This simple computational exercise can confirm or dismiss the experimental 
conjecture regarding whether the mechanochemical strain significantly induces 
regulation on the nucleotide binding site and also if it occurs in the L. 
\\

{\bf Catalytic core of the leading head is less native-like on the MT: }
Since the nucleotide binding and release dynamics is sensitively controlled by the kinesin structure, 
we assume that the nucleotide molecule has an optimal binding affinity to the kinesin motor domain in the native structure. 
For function there is a need to understand how the native structure of the 
kinesin motor domain is perturbed 
under the different topological constraints imposed on the dimeric kinesin configuration by interacting with the MT. 
The equilibrium ensemble of the structures shows that the neck-linker is in the 
docked state for the T but undocked for L. 
In comparison to the native structure, the overall shape of the 
nucleotide binding pocket in the T is more preserved. 
As long as the MT constrains the two heads 8-nm apart, this configuration is 
dominant in the thermal ensemble (see Fig.3-A).

{\it Global shape comparison :}
There are in principle multiple ways to quantitatively compare the two motor domain structures. 
To assess the structural differences, the radius of gyration 
($R_g^2=1/2N^2\sum_{i,j}(\vec{R}_i-\vec{R}_j)^2$)
of the two motor domain structures from the equilibrium ensemble are computed (see Fig.3-B). 
Because the neck-linker and the neck-helix adopt different configurations 
relative to the motor domain in each monomer, we perform a $R_g$ analysis for the motor domains only (residue 2-324). 
The $R_g$ distributions show that the L is slightly 
bigger than the T both in the size ($(\langle R_g\rangle(L)-\langle R_g\rangle(T))\sim 0.4$ \AA) and 
in the dispersion ($(\sigma_L-\sigma_T)\sim 0.05$ \AA). 
Meanwhile the $R_g$ for the native state (3kin) is $R_g^{native}=19.4$ \AA. 
Clearly, the sizes of both of the heads in the thermal ensemble 
are expanded at $T=300K$ as compared to the native structure. 
The size alone does not tell much 
about the difference between the structures. 

The RMSD relative to the native structure and 
between the two motor domains (residues 2-324) computed over 
the equilibrium ensemble gives 
$\mathrm{RMSD}(T|native)=2.0$ \AA, $\mathrm{RMSD}(L|native)=9.4$ \AA, $\mathrm{RMSD}(T|L)=9.6$ \AA, where $\mathrm{RMSD}(X|Y)$ is 
the RMSD between conformations X and Y. 
If the $\alpha6$ helix is excluded from the RMSD calculation of motor domain 
(residues 2-315), then 
$\mathrm{RMSD}(T|native)\approx 1.8$ \AA, $\mathrm{RMSD}(L|native)\approx 3.8$ \AA, and $\mathrm{RMSD}(T|L)\approx 3.9$ \AA. 
The RMSD analysis shows that the 
$\alpha6$ helix significantly contributes more for the deviation of the leading  head from its native state than the T. 

Additional detailed comparisons with respect to the native state can be 
made using the structural overlap function of $i-j$ pair, $\chi_{ij}$, which is defined as $\chi_{ij}=\langle \delta(R_{ij}-R^o_{ij}) \rangle$
where $\delta(R_{ij}-R^o_{ij})=1$ if $|R_{ij}-R^o_{ij}|<r_{tol}$, otherwise $\delta(R_{ij}-R^o_{ij})=0$. 
$R_{ij}^o$ is the distance of $i-j$ pair in native state, where $r_{tol}=1$ \AA. 
By setting $R^o_{ij}$ values identical in both heads (i.e., both heads have the same native state), we compute the $\chi_{ij}$ values for the trailing and the leading heads, respectively.  
The relative difference of the $\chi_{ij}$ value between T and L, $X_{ij}$ is defined by
\begin{equation}
X_{ij}=\left\{ \begin{array}{cc}
     \frac{\chi_{ij}(T)-\chi_{ij}(L)}{\chi_{ij}(T)}& \mbox{$(\chi_{ij}(T)\neq 0)$}\\
     0& \mbox{$(\chi_{ij}(T)= 0)$}\end{array}\right.
\label{eqn:Chi}
\end{equation}
which quantitatively measures the structural difference of the two heads. 
Based on the $X_{ij}$ value (Fig.3-C), the distances between the MT binding 
motif of the T (L11, L12, $\alpha4$, $\alpha 5$) and other secondary structure units ($\beta 1$, $\alpha 0$, $\beta 2$, $\alpha 1$, $\alpha 2$, $\alpha 6$) are 50 \% more native-like than in the L. 

{\it Conserved native contacts in trailing head reveals the strain propagation pathway in leading head: }  
A direct measure of similarity to the native structure is 
the fraction of native contacts preserved in the thermal ensemble \cite{Cho06PNAS}. 
Since we assume that ATP affinity is optimized in the native state, 
we can readily assess the quality of the structure using this measure. 
We quantify the nativeness of a pair using 
$q_{ij}(\xi)=\langle\Theta(R_c-R_{ij})\Delta_{ij}\rangle$,
where $\Delta_{ij}=1$ if $i$, $j$ residues are in contact at the native state ($R^o_{ij}<R_c^K=8$ \AA), and $\Delta_{ij}=0$ otherwise. $q_{ij}(\xi)$ (with $\xi=T$ or $L$) is obtained by averaging over the thermal ensemble. 
When $q_{ij}(\xi)$ is averaged over all the native pairs, the average fraction of native contacts, 
$\langle Q\rangle$ is calculated as $\langle Q\rangle(\xi) =1/N_Q\sum^{N_Q}_{i<j}q_{ij}(\xi)$ where $N_Q$ is the total number of native pairs. 
For the T and L conformations, 
$\langle Q\rangle(T)=0.86$ and $\langle Q\rangle(L)=0.82$, respectively. 
The relative difference of native contacts between the two kinesin heads at the pair level, $Q_{ij}$, is quantified similarly to Eq.\ref{eqn:Chi} as 
\begin{equation}
Q_{ij}=\left\{ \begin{array}{cc}
     \frac{q_{ij}(T)-q_{ij}(L)}{q_{ij}(T)}& \mbox{$(q_{ij}(T)\neq 0)$}\\
     0& \mbox{$(q_{ij}(T)= 0)$}\end{array}\right.
\label{eqn:Phi}
\end{equation}
In Fig.4, $Q_{ij}$ is color-coded based on its value. 
As expected from the equilibrium ensemble, 
conspicuous differences are found around the structural motifs 
having direct contacts with neck-linker, giving $Q_{ij}\gtrsim 0.5$. 
Quantitative inspection of the other contacts is illustrated 
in the structure. 
We color the kinesin head structure based on the $Q_{ij}$ value. 
The residue pairs are colored in magenta if $0.2<Q_{ij}<0.5$, red if $Q_{ij}>0.5$, where the positive $Q_{ij}$ signifies that the native contacts in trailing head are more intact. 
The residue pairs are colored in light-blue if $-0.5<Q_{ij}<-0.2$, blue if $Q_{ij}<-0.5$. 
More intact contacts, when the trailing and the leading head are compared, are visualized by yellow line in Fig.5-B. 
Our analysis not only shows that there is higher probability of the formation of native contacts present in the T in comparison to the L, but also suggests how the tension is propagated towards the 
nucleotide binding site to disrupt the nativeness of the nucleotide binding pocket in the leading head. 
As expected, a dense network of intact contacts are found between the 
neck-linker ($\beta 10$) and the neck-linker binding motif ($\beta 7$). 
This network continues along the $\alpha 6$ helix, perturbing $L2$, $\beta1$, $\alpha4$, and finally reaches the nucleotide binding site (see SI Fig.6 for the nomencaltures of the secondary structures). 
It is surprising that the disruptions of native contacts are found particularly 
in the nucleotide binding site, which is believed to be the trigger point for the 
allosteric transition. 
All the important nucleotide binding motifs (P-loop, switch-1, switch-2, and N4) are recognized by our simulational analysis using a nonlinear-Hamiltonian 
(see Supporting Information for the comparison with linear-harmonic potential represented as Gaussian network model).
\\

{\bf Estimate of the tension in the neck-linker: }
The deformation of the leading motor domain is caused by the internal 
tension in the neck linker. 
The tension on the neck linker is estimated using the force ($f$) versus extension ($x$) relationship of a worm-like chain model \cite{MarkoMacro96}, 
\begin{equation}
f=\frac{k_BT}{l_p}\left[\frac{1}{4(1-\frac{x}{L})^2}+\frac{x}{L}-\frac{1}{4}\right], 
\label{eqn:WLC}
\end{equation}
where $l_p$ is the persistence length of the polymer and $L$ is the contour length. 
$L\approx 5.7 nm$ for the 15-amino acid neck-linker (residue from 324 to 338) $(= 15\times 0.38 nm)$, and in the equilibrium ensemble of structures, $x\approx 3.1\pm 0.8$ $nm$. 
Assuming that $l_p\approx(0.4-0.5)$ $nm$ \cite{EatonPNAS05} for this segment, 
we estimate a tension $f=12-15$ $pN$. 
By integrating Eq.\ref{eqn:WLC} for $(0-3.1)$ $nm$. 
The tensional energy stored in the neck-linker is obtained, which is 17 $pN\cdot nm\approx 4k_BT$. About 20\% of the ATP hydrolysis energy ($\sim 25 k_BT$) is 
stored in the neck-linker and directly perturbs the nucleotide binding 
site of the L, 
whereas mechanical action to the T is dissipated 
through the dense network of contacts formed between 
the neck-linker ($\beta10$) and the neck-linker binding site ($\beta7$). 

For a given extension $x(<L)$, when the length of the neck-linker is varied by $\delta L$, the variation in the length of the neck-linker can affect the effective tension as 
$f(L+\delta L)-f(L)\approx -k_BT/l_p[1/\{4(1-x/L)^2\}\times(2x/L)/(1-x/L)+x/L](\delta L/L)$.
For a given $\delta L/L$, the resulting tension change may be significant depending on the value of the extension $x$. 
Experimentally, the kinesin dynamics has recently 
been studied by varying the linker length by introducing a spacer 
composed of amino acids \cite{Hackney03Biochem}. Since $\delta L=0.38$ $nm$ 
for the insertion of a single amino acid, the lengthening of the linker leads to a reduction of the tension by $\Delta f\approx -2 pN$. In light of the force values controlling the kinesin dynamics in LOT experiments, which is $(0-7)$ $pN$ and 7 $pN$ is the stall force, a value of $\Delta f\approx -2pN$ can be significant. 
According to Hackney \emph{et. al.}'s experimental analysis the processivity is reduced by $\approx 2$-fold when a single 
amino acid is inserted or deleted, and 6 or 12 additional amino acids resulted in to a 3$-$4 fold reduction in the kinetic processivity \cite{Hackney03Biochem}. 
\\

\section{Conclusions}
Because of the size and time scale spanned by a typical molecular motor as well as the lack 
of crystal structures, theoretical study based on the structure is not a common approach 
such as master equation descriptions \cite{TeradaPNAS02,Fisher01PNAS,Fisher05PNAS} or Brownian ratchet models \cite{Reimann02PR}.
Knowledge of structural details, however, is the key ingredient to understand the mechanochemistry of molecular motors. 
In the present study, we propose computational strategies to resolve this problem in kinesin dynamics, particularly the 
nucleotide binding regulation mechanism between the two motor domains. 
By building a kinesin model on the tubulin filament, we explicitly identified the effect of internal tension
($f=12-15$ $pN$) on the front kinesin head domain and showed that the tension propagation to the leading head provokes 
the switch-related motifs (P-loop, switch-1, switch-2) in the nucleotide binding pocket. 
Assuming ``the nativeness as a criterion for the optimal nucleotide 
binding condition'', 
we concluded that the nucleotide binding pocket in the leading head is not favorable (or partially unfolded or cracked \cite{Miyashita03PNAS}) 
for the nucleotide binding while the trailing head is bound to the MT.
This conclusion explains the recent real-time single molecule traces of kinesin 
generated by Guydosh and Block \cite{BlockPNAS06}.
The reduction of ligand affinity of the leading head due to the rearward 
tension benefits the high processivity in two ways. 
First, the premature ATP binding is inhibited before the chemical reaction (ATP hydrolysis, $P_i$ release) in the trailing head is completed, during which the trailing head is tightly bound to the MT and the tension is built on the neck linker. 
Second, the release of ADP is facilitated in the leading head, 
which accelerates step $(ii')\rightarrow (iii)$ in Fig.1.
The latter point is consistent with Uemura and Ishiwata's experimental observation of a seven-fold increase of the ADP dissociation constant from the monomeric kinesin head in the presence of rearward loading \cite{Uemura03NSB}. 

It is noteworthy that global measures characterizing the two head domains do not distinguish the qualitative differences 
between the two heads. Major changes in the strained region such as the kinesin neck-linker may be associated to minor global 
differences in the motor domain. 
Exactly how these minor global differences that are localized in a small volume is amplified 
in the process of allosteric transition is the another key issue in understanding molecular motors.  

The internal tension regulates the interaction between the kinesin and the 
nucleotide. Conversely, the interaction between kinesin, nucleotide, and MT also switches on and off the internal tension. 
For conventional kinesin, mechanical and chemical mechanism are closely correlated, producing 
a remarkable processivity of the kinesin movement on the MT.

\section{Methods}
The simulations were performed using two classes of energy function. 
One is the standard structure based (SB) potential \cite{ClementiJMB00} and the other is the self-organized polymer (SOP) potential \cite{HyeonBJ07,HyeonSTRUCTURE06,Hyeon06PNAS} (see Supporting Information and SI Figs. 7-8). 
For these two different topology-based potential 
functions, we obtained qualitatively identical results. 
Results from the SB potential are presented in Figs.3-5 while results from the SOP potential are in SI Fig.9 in Supporting materials. 
This suggests that the results are robust as long as the information of 
native topology with a nonlinear form of energy potential 
is used as an input. 
\\

{\bf Energy function$-$Structure based potential: }
Using the structure in Fig.2-C as a starting structure, 
we simulated and sampled an ensemble of two-headed kinesin configurations on the MT. 
We performed the Langevin simulations of the SB model \cite{ClementiJMB00} whose equation of the motion of each interaction center is integrated by a Verlet algorithm.
The energy potential is given as 
\begin{align}
&H(\{\vec{r}_i\})=\{H^K_{bond}+H^K_{nb}\}+H^{K-tub}_{nb}\nonumber\\
&=\sum_{i=1}^{N_K-1}\frac{K_r}{2}(r_{i,i+1}-r^o_{i,i+1})^2+\sum_{i=1}^{N_K-2}\frac{K_{\theta}}{2}(\theta_i-\theta_i^o)^2+\sum_{i=1}^{N_K-4}\sum_{n=1,3}K_{\phi}^{(n)}(1-\cos{[n(\phi_i-\phi_i^o)]})\nonumber\\
&+\sum_{i=1}^{N_K-3}\sum_{j=i+3}^{N_K}\left[\epsilon_h\left(\left(\frac{r^o_{ij}}{r_{ij}}\right)^{12}-2\left(\frac{r^o_{ij}}{r_{ij}}\right)^6\right)\Delta_{ij}+\epsilon_l\left(\frac{\sigma}{r_{ij}}\right)^{12}(1-\Delta_{ij})\right]\nonumber\\
&+\sum_{i=1}^{N_K}\sum_{k=1}^{N_{tub}}\left[\epsilon_h\left(\left(\frac{r^o_{ik}}{r_{ik}}\right)^{12}-2\left(\frac{r^o_{ik}}{r_{ik}}\right)^6\right)\Delta^*_{ik}+\epsilon_l\left(\frac{\sigma}{r_{ik}}\right)^{12}(1-\Delta^*_{ik})\right],
\end{align}
where the energy Hamiltonian is divided into intramolecular interactions for 
the kinesin molecule and intermolecular interactions at the kinesin-MT interface. 
The superscripts $K$ and $K-tub$ denote the kinesin and the kinesin-tubulin interaction, respectively. 
Because our focus is on the kinesin dynamics, we fixed the coordinates of the MT in space. 
Since the length scale of the kinesin geometry is small ($<10nm$) compared to that of MT (diameter $\sim 24 nm$, persistence length $\sim 1 mm$), 
the explicit computation of the dynamics of the entire MT structure, in which the 13 protofilaments constitute the cylindrical geometry, should not qualitatively change our conclusions. 
The first and the second term define the backbone interactions. 
The bond distance $r_{i,i+1}$ between the neighboring residues $i$ and $i+1$ are harmonically constrained with respect to the bond distance in native state $r_{i,i+1}^o$ with a strength $K_r=20$ $kcal/(mol\cdot \AA^2)$. 
In the second term, the angle $\theta$ is formed between residues $i$, $i+1$, and $i+2$ with $K_{\theta}=20$ $kcal/(mol\times rad^2)$. $\theta_i^o$ is the angle of the native state. 
The third term is the dihedral angle potential with $K_{\phi}^{(1)}=1.0 kcal/mol$ and $K_{\phi}^{(3)}=0.5 kcal/mol$ that describes the ease of rotation around the angle formed between successive residues from $i$ to $i+3$ along the backbone. 
The Lennard-Jones potential is used to account for the interactions that 
stabilize the native topology. 
A native contact is defined from the pair of interaction centers whose distance 
is less than $R_c^K=8$ \AA\ in native state for $|i-j|>2$. 
If $i$ and $j$ residues are in contact in the native state, $\Delta_{ij}=1$, otherwise $\Delta_{ij}=0$. 
Aided by $\Delta_{ij}$ we assign stabilizing potential for native pairs and repulsive potential for non-native pairs. 
We assign $\epsilon_h=1.8$ $kcal/mol$ for the intra and inter neck-helix (residue$>338$) interactions to secure the coiled-coil association between the neck-helices. 
For other kinesin residue-residue interactions, we set $\epsilon_h=\epsilon_l=1.0$ $kcal/mol$ regardless of the sequence identity. 
The parameters determining the native topology $r_{ij}^o$ and $\Delta_{ij}$ are determined from the crystal structure of human kinesin (PDB id : 3kin) and incorporated to the trailing kinesin (T) and the coiled-coil whose structure is shown in Fig.2-C. 
To constitute an identical fold condition,  
we transferred topological information in trailing head (T) to the leading head (L) by substituting $r_{ij}^o$, $\theta_i^o$, $\phi_i^o$ and $\Delta_{ij}$ 
from T to L, i.e., 
$r^o_{ij}(L)=r^o_{ij}(T)$, $\theta_i^o(L)=\theta_i^o(T)$, $\phi_i^o(L)=\phi_i^o(T)$ and $\Delta_{ij}(L)=\Delta_{ij}(T)$ for all $i$ and $j$. 
The kinesin-tubulin interaction parameters ($\epsilon_h$, $\epsilon_l$, $r_{ik}^o$, and $\Delta_{ik}^*$) are similarly defined as kinesin intramolecular interaction parameters except for the slightly larger native contact distance ($R_c^{K-tub}=10$ \AA).
The parameters $r_{ik}^o$ and $\Delta_{ik}^*$ defining the interface topology between the trailing kinesin head (T) and the tubulin are also transfered 
to the interface topology between the leading head L and the next tubulin binding site. 
\\

{\bf Simulations :} 
The initial structure, whose two heads are constrained to be oriented on the tubulin binding sites, is relaxed 
under the SB or SOP-Hamiltonian and subsequently the equilibrium ensemble of the structures is collected from the low friction Langevin dynamics simulations at $T=300 K$. 
The position of the interaction center is integrated using 
\begin{equation}
m\ddot{\vec{r}}=-\zeta\dot{\vec{r}}-\frac{\partial H}{\partial\vec{r}}+\vec{\Gamma}
\end{equation}
where $\zeta$ is friction coefficient, $-\frac{\partial H}{\partial\vec{r}}$ is the conformation force, and $\vec{\Gamma}$ is the random force satisfying $\langle\vec{\Gamma}(t)\cdot\vec{\Gamma}(t')\rangle=\frac{6\zeta k_BT}{h}\delta(t-t')$ where the integration time ($h$) is discretized. 
In low friction Langevin dynamics, natural time is given by $\tau_L=(ma^2/\epsilon_h)^{1/2}$. 
We chose $\zeta=0.05\tau_L^{-1}$ and $h=0.0025\tau_L$. 
Low friction is deliberately chosen for the purpose of effectively sampling the conformational space \cite{HoneycuttBP92}.
Under such conditions the resulting dynamics as a function of time step should not be taken parallel to the real time dynamics. 
To produce an overdamped dynamics it is essential to integrate the motion by neglecting 
the inertial term as well as choosing a high friction coefficient that amounts to the water viscosity ($\sim 1cP$). 
\\

\section{Acknowledgements}
We are grateful to Paul Whitford and Sam Cho for critically reading the manuscript. 
This work was funded by the NSF Grant 0543906 and by the NSF-sponsored Center for Theoretical Biological Physics (Grants PHY-0216576 and 0225630). 

\clearpage

\section{Figure Legends}

{{\bf Figure} 1: }
Mechanochemical cycle of kinesin. 
The subscripts $X$ and $Y$ refer to each of the kinesin head, $E$ denotes 
the empty head, and (:MT) is appended if the head is strongly bound to 
the MT.

{{\bf Figure} 2:} 
Procedure to construct the two-headed kinesin/MT-protofilament model. 
{\bf A.} Three structures from protein data bank (PDB) are used. 
1. Single headed kinesin (KIF1A) bound on tubulin (PDB id: 1ia0). 
2. Two-headed kinesin (PDB id: 3kin).
3. Two consecutive tubulin complexed to the stathmin-like domain (PDB id: 1ffx). 
We overlaped the chain (A), (B) (blue in figure) of 3kin onto the chain K of 1ia0 and the chain A of 1ffx ($\alpha$-domain) onto the chain (B) of 1ia0, which leads to the structure in {\bf B}. 
The structural homology ($C_{\alpha}$ backbone RMSD=1.6\AA) between KIF1A and a head of the two-headed kinesin are sufficient that one of the kinesin heads 
fits to the tubulin binding site. 
While the sequence difference between KIF1A and the conventional kinesin (sequence identity $\sim 45$\%) may affect the strength of 
interactions between kinesin and tubulin, leading to a different binding affinity of conventional kinesin from KIF1A, 
we assume that the binding orientation of 
the two-headed kinesin is similar to that of KIF1A on the tubulin. 
After the structure overlap, the chain (C), (D) of 3kin is internally 
rotated around a few positions in the neck linker (324-338) 
until the chains C, D are placed in the 
vicinity of binding site of tubulin that is designed to be identical to the interface between 
the kinesin rear head and the tubulin. 
We performed the simulation to relax the kinesin structure on the MT and obtained the structure shown in {\bf C}. 

{{\bf Figure} 3: }
The ensemble of structures and structural comparisons between 
two heads using $R_g$ and $X_{ij}$.
{\bf A.} The thermal ensemble of structures is illustrated using the multiple structures obtained during the simulations. 
Different colors are used to distinguish the motor domain (residues 2$-$323) from the neck-linker and the tail part (residues 324$-$370). 
Substantial variations of the neck-linker/tail position in ensemble 
show its flexibility. 
The nucleotide binding sites in the L and T are indicated by the arrows. 
On the upper-right corner of the panel are two crystal structures of kinesin (3kin). 
One is the view from the top (left) and the other is the view towards the 
nucleotide binding pocket (right). 
When compared with the top view of 3kin crystal structure, it is 
visually clear that the nucleotide binding pocket is more intact in the T.   
{\bf B.} Analysis of kinesin motor domains (2-324) using the radius of gyration ($R_g$). 
Histograms of $R_g$ collected over the ensemble are fit to a Gaussian distribution. 
For the trailing head (red), $\langle R_g\rangle_T=20.3$ \AA, $\sigma_T=0.13$ \AA. 
For the leading head (blue), $\langle R_g\rangle_L=20.7$ \AA, $\sigma_L=0.18$ \AA. 
{\bf C.} Analysis using the structure overlap function. 
$X_{ij}$ value is color-coded on the right (see the text for the definition of $X_{ij}$). 

{{\bf Figure} 4:}
Comparison of two heads using the fraction of native contact.
{\bf A}. The average contact map for kinesin (left). 
The relative difference of average contact map between trailing head and leading head with respect to trailing head is shown on the right. 
The red arrows mark the set of contact pairs whose $Q_{ij}$ value is greater than 0.5 (50\%). {\bf B}. The result of $Q_{ij}$ in {\bf A} is redrawn using 3-D plot for clarity.

{{\bf Figure} 5:}
The structure of kinesin on MT colored based on the protocol discussed in the main text. 
{\bf A.}
The residues colored in red and magenta on the trailing head (structure on the left) are those maintaining more contact than the residues in the leading head (structure on the right). 
Some of the residues in colored region of the trailing head are involved with the nucleotide binding site. {\bf B.} The enlarged view in cartoon representation 
from neck-linker docking site (left) and from ligand binding site (right). 
Sphere representations in orange are ADP molecule. The network of contact pairs 
are depicted in yellow lines. 

\clearpage

\begin{figure}[ht]
\includegraphics[width=6.0in]{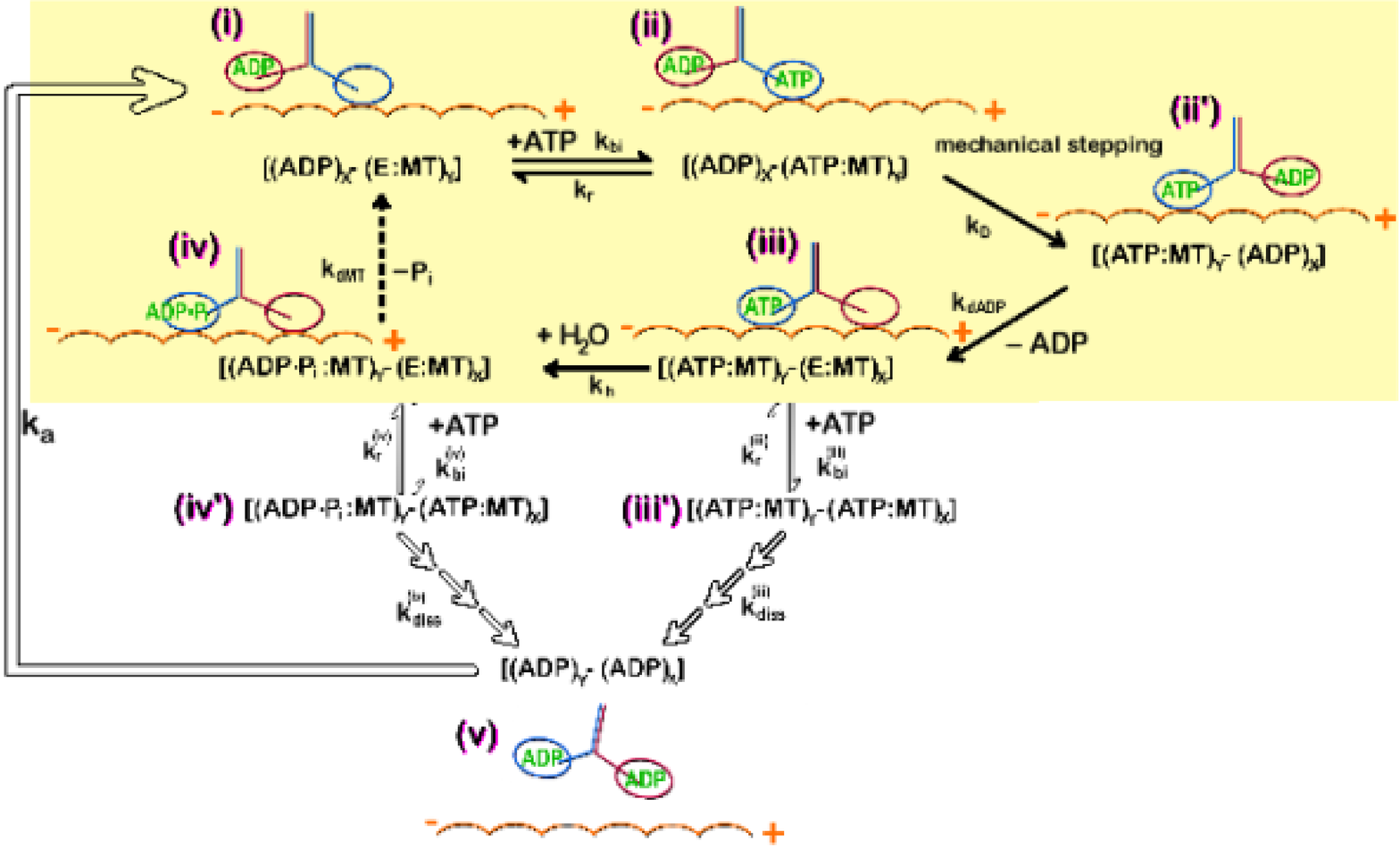}
\caption{\label{Fig1.fig}}
\end{figure}
\clearpage
\begin{figure}[ht]
\includegraphics[width=5.00in]{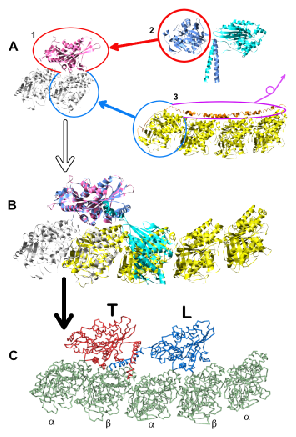}
\caption{\label{Fig2.fig}}
\end{figure}
\clearpage
\begin{turnpage}
\begin{figure}[ht]
\includegraphics[width=8.0in]{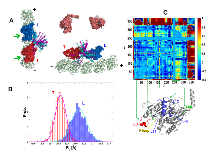}
\caption{\label{Fig3.fig}}
\end{figure}
\end{turnpage}
\clearpage
\begin{turnpage}
\begin{figure}[ht]
\includegraphics[width=9.00in]{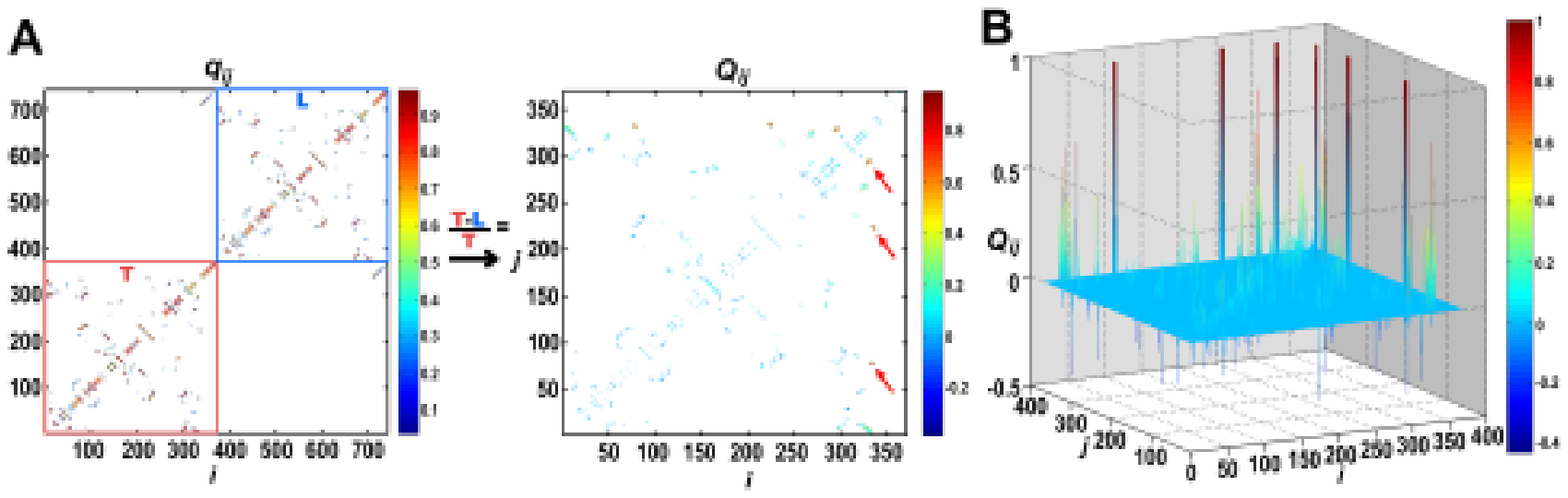}
\caption{\label{Fig4.fig}}
\end{figure}
\end{turnpage}
\clearpage
\begin{figure}[ht]
\includegraphics[width=6.00in]{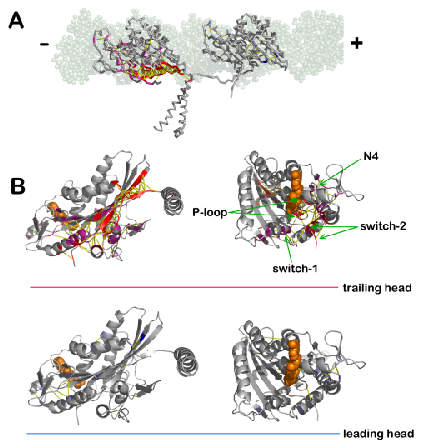}
\caption{\label{Fig5.fig}}
\end{figure}
\clearpage

{\bf\Large Supporting Information}\\

{\bf Structural Details of Kinesin.} 
Detailed knowledge of the structure is fundamental to understand the working mechanism of a biological nanomachine. 
A monomer of the conventional kinesin is structurally categorized into three parts : The head (or motor domain) (residues 2$-$323), the neck linker (residues 324$-$338 : $\beta 9$, $\beta 10$), and the neck-helix (residue 339$-$ : $\alpha 7$). 
The head region is composed of eight $\beta$-strands flanked with three $\alpha$-helices [$\alpha 1$, $\alpha 2$, $\alpha 3$ on one side (Fig. 6A), and 
$\alpha 4$, $\alpha 5$, $\alpha 6$ on the other side (Fig. 6B)] on each side of the $\beta$-sheet. 
One side contains the binding motif ($\alpha 4$, $\alpha5$, $\alpha 6$, L8, L11, L12) for the microtubule (Fig. 6A) 
and the other side provides a nucleotide binding pocket (Fig. 6D). 
The nucleotide binding site in the kinesin head region is structurally homologous 
across the motor protein and the G-protein superfamilies. 
The structural motifs around the nucleotide binding site, such as the P-loop (N1) (86$-$93), switch-1 (N2) (199$-$204), switch-2 (N3) (232$-$237), N4 (14-17) are accordingly designated \cite{Mandelkow97BC,Sack99EJB}. 
The crystal structures with the different nucleotide states suggest that the presence or 
the absence of $\gamma$-$P_i$ is sensed by these motifs and that the structural 
changes of the motifs are related to the allosteric transitions. 
The ordered state of the neck-linker, which is extended from the N terminus of kinesin head and is composed of two beta strands ($\beta 9$ and $\beta 10$), 
forms contacts with the N-terminal region of the $\beta 7$-strand. 
Further extension from the neck-linker leads to the neck-helix ($\alpha 7$-helix), through which two monomers form a dimeric complex. \\

{\bf Computations of Residue Displacement Cross-Correlation Using Elastic Gaussian Network Model and Simulation under SB Potential. }
For given coordinates of a complex three-dimensional structure, 
the dynamical property of an object can be extracted at the zeroth order by 
investigating its topology. 
The Gaussian network model (GNM) is the simplest possible method to study large biomolecules using the corresponding minimal topology \cite{TirionPRL96,BaharPRL97,BrooksPNAS00}. 
The GNM views the biomolecular construct as a collection of beads connected with harmonic springs with strength $\gamma$. 
The connectivity between the beads is purely determined by the cut-off distance parameter $R_C$. 
For a pair of beads $i$ and $j$, whose distance $R_{ij}(=|\vec{R}_i-\vec{R}_j|)$ satisfies $R_{ij}<R_C$, the harmonic potential constrains the position of beads via
\begin{equation}
H=\sum_{i<j}\frac{\gamma}{2}(\vec{R}_{ij}-\vec{R}^o_{ij})^2\Theta(R_c-R_{ij})=\frac{\gamma}{2}\delta\mathbf{R}^T\cdot\Gamma\cdot\delta\mathbf{R},
\tag{6}
\end{equation}
where $\Theta(\ldots)$ is the Heaviside function, $\delta\mathbf{R}^T=(\delta\vec{R}_1,\ldots,\delta\vec{R}_N)$ with $\delta\vec{R}_i=\vec{R}_i-\vec{R}_i^o$,  
and $\Gamma_{ij}=\frac{1}{2}\frac{\partial^2 H}{\partial\delta R_i\partial\delta R_j}$ is the $(i,j)$ element of Kirchhoff matrix $\Gamma$.
Use of the harmonic potential amounts to the expansion of the potential $H(\{\mathrm{R}\})$ at the potential minimum $\{\mathbf{R}^o\}$ as 
$H(\{\mathbf{R}\})=H(\{\mathbf{R^o}\})+\frac{1}{2}\delta\mathbf{R}^T\frac{\partial^2H}{\partial\delta\mathbf{R}\partial\delta\mathbf{R}}\delta\mathbf{R}+\cdots$.
Since the partition function of GNM is given by 
$Z_N=\int D[\delta\mathbf{R}]e^{-\beta H}=\left[\det\left(\frac{\gamma\Gamma}{2\pi k_BT}\right)\right]^{-3/2}$,
the correlation between the spatial fluctuation of two residues is expressed using  
the inverse of Kirchhoff matrix, 
\begin{equation}
\langle \delta R_i\cdot\delta R_j\rangle=-\frac{2 k_BT}{\gamma}\frac{\partial\log{Z_N}}{\partial \Gamma_{ij}}=\frac{3k_BT}{\gamma}\left(\Gamma^{-1}\right)_{ij}.
\label{eqn:Z_N}
\tag{7}
\end{equation}
For $i=j$, the mean square displacement of the $i^{th}$ residue, $\langle \delta R_i^2\rangle$, corresponds to the B-factor (Debye-Waller temperature factor) as $B_i=\frac{8\pi^2}{3}\langle \delta R_i^2\rangle$. 
A comparison between the B-factor and mean square displacement (MSD) from the 
GNM determines the effective strength of the harmonic potential that stabilizes 
the structure. 
Note that the quality of the MSD in GNM is solely controlled by the $R_C$ value, thus we scaled the MSD with $3k_BT/\gamma$ for GNM analysis. 

We applied the GNM analysis with $R_C=8$ \AA\ on the two-headed kinesin whose both heads fit to the adjacent tubulin binding site, and then computed the cross-correlation matrix as shown in Fig.7A. 
The cross-correlation value $C_{ij}$, $\langle\delta R_i\cdot\delta R_j\rangle$ scaled by $3k_BT/\gamma$, shows that except for the neck-helix region the amplitude of 
correlation in leading head is always larger than that of the trailing head. 
This is expected since the neck-linker of the leading kinesin is detached from the motor domain. 
The residues in the network with less coordination number are subject to a larger fluctuation. 
The relative difference of the cross-correlation between the leading and the trailing kinesin using 
\begin{equation}
\delta_{ij}=\frac{C_{ij}(i,j\in L)-C_{ij}(i,j\in T)}{C_{ij}(i,j\in L)}
\tag{8}
\end{equation}
is illustrated in Fig. 7B.
Fig. 7C shows the auto-correlations (or mean square displacement), which are the diagonal elements of the $C_{ij}$ matrix. 

GNM analysis is useful in analyzing the fluctuation dynamics of the stable structure at the residue level in the basin of attraction where the basin is modeled as a quadratic potential. 
However, the expansion of the potential minima up to the quadratic 
term is justified only if the fluctuation $\delta\mathrm{R}$ is small. 
The amplitude of fluctuation in biological systems at physiological temperatures ($T\sim 310 K$) is most likely to exceed the limit beyond which nonlinear response is no longer negligible. 
In order to take this effect into account, the Hamiltonian should be expanded beyond the linear response regime. 
This procedure indeed reverses the simple idea that Tirion \cite{TirionPRL96} and Bahar et al. \cite{BaharPRL97} have proposed in the context of GNM analysis. 
However, \emph{minimal} inclusion of the nonlinear term can be useful by increasing 
the susceptibility of the structure. 
Once the nonlinear term is included, a simple analytical expression such as Eq.\ref{eqn:Z_N} is not available. 
Thus, we resort to the simulations. 

The analytically obtained quantities, $\langle\delta R_i\cdot\delta R_j\rangle$, $\delta_{ij}$ in Fig. 7
can also be calculated over the thermal ensemble of structures obtained from 
simulations using a nonlinear-Hamiltonian (see Fig. 8). 
The first conclusion drawn from the simulational analysis is similar to the GNM in that 
the leading head experiences larger fluctuations. 
Secondly, the position and relative amplitude of the MSD peaks, 
reproduced using the simulation results, shows a good agreement with GNM results. 
However, the direct comparison of $C_{ij}$ (or $\delta_{ij}$) between Fig. 7 and Fig. 8 shows that the simulation results from the nonlinear-Hamiltonian display a more sensitive pattern 
of cross-correlations. 
The pronounced amplitude of $C_{ij}$ (or $\delta_{ij}$) 
suggests a strong spatial correlation between residues $i$ and $j$. 
\\

{\bf Alternative Energy Function : SOP Potential.} 
An alternative potential function for the SB potential used in the main text is the self-organized polymer (SOP) potential that was recently adopted for simulations of the mechanical unfolding of large molecules of RNA and proteins \cite{HyeonSTRUCTURE06,HyeonBJ07} as well as the allosteric dynamics of GroEL \cite{Hyeon06PNAS}. 
The energy Hamiltonian is defined as 
\begin{align}
H(\{\vec{r}_i\})&=\{H^K_{FENE}+H^K_{nb}\}+H^{K-tub}_{nb}\nonumber\\
&=-\sum_{i=1}^{N_K-1}\frac{k}{2}R_0^2\log({1-\frac{(r_{i,i+1}-r_{i,i+1}^o)^2}{R_0
^2}})\nonumber\\
&+\sum_{i=1}^{N_K-3}\sum_{j=i+3}^{N_K}\epsilon_h\left[\left(\frac{r^o_{ij}}{r_{ij}}\right)^{12}-2\left(\frac{r^o_{ij}}{r_{ij}}\right)^6\right]\Delta_{ij}\nonumber\\
&+\sum_{i=1}^{N_K-2}\epsilon_l\left(\frac{\sigma}{r_{i,i+2}}\right)^6+\sum_{i=1}^{N_K-3}\sum_{j=i+3}^{N_K}\epsilon_l\left(\frac{\sigma}{r_{ij}}\right)^6(1-\Delta_{ij})\nonumber\\
&+\sum_{i=1}^{N_K}\sum_{k=1}^{N_{tub}}\left[\epsilon_h\left(\left(\frac{r^o_{ik}}{r_{ik}}\right)^{12}-2\left(\frac{r^o_{ik}}{r_{ik}}\right)^6\right)\Delta^*_{ik}+\epsilon_l\left(\frac{\sigma}{r_{ik}}\right)^6(1-\Delta^*_{ik})\right]. 
\label{eq:SOP}
\tag{9}
\end{align}
The first term is for the chain connectivity of the kinesin molecule. 
The finite extensible nonlinear elastic (FENE) potential \cite{KremerJCP90} is used
with $k=20kcal/(mol\cdot$\AA$^2)$, $R_0=2$ \AA,
and $r_{i,i+1}$ is the distance between neighboring interaction centers
$i$ and $i+1$. 
The Lennard-Jones potential interactions stabilize
the native topology.
A native contact is defined as the pair of interaction centers whose distance is
less than $R^K_C=8$ \AA\ in native state for $|i-j|>2$.
If $i$ and $j$ sites are in contact in the native state, $\Delta_{ij}=1$, otherwise
$\Delta_{ij}=0$.
We used $\epsilon_h=1.8$ $kcal/mol$ in the native pairs, and $\epsilon_l=1$ $kcal/mol$ for non-native pairs.
To ensure the non-crossing of the chain, we used a $6^{th}$ power potential in the repulsion terms and set $\sigma=3.8$\AA,
which is typical $C_{\alpha}-C_{\alpha}$ distance. 
The parameters determining the native topology, $r^o_{ij}$ and $\Delta_{ij}$, 
are adopted from the trailing kinesin (X) whose structure is shown in Fig. 2{\it C}.
We transferred the topological information in the trailing head (T) to the leading head (L) by substituting 
$r^o_{ij}$ and $\Delta_{ij}$ from the T to L. 
Kinesin-tubulin interaction energies are similarly defined as kinesin intramolecular interaction energies with slightly different native contact distances. 
We set the cut-off distance for the 
native interactions between the kinesin and the 
tubulin as $R_C^{K-tub}=10$ \AA. 
The parameters, $r_{ik}^o$ and $\Delta^*_{ik}$, 
defining the interface topology between the kinesin head T and the tubulin is 
transfered to the kinesin head L and the next tubulin binding site. 
Using the SOP potential, we obtained qualitatively identical results 
as those obtained from the SB potential. 
The nucleotide binding pocket of the front head is disrupted in the dimeric 
kinesin configuration whose both heads are bound to the tubulin binding sites.
The figures corresponding to Fig. 3{\it A} and {\it C} and Fig. 5 are regenerated using SOP model in Fig. 9.
\\

{\bf Master Equations for the Mechanochemical Cycle of Kinesin Described in Fig. 1. }
In the limit when the dissociation of dimeric kinesin from the microtubule is 
negligible, the kinetic equation describing the dynamic cycle shown in Fig. 1 is written as 
\begin{align}
\frac{dP_{(i)}}{dt}&=-k_{bi}[ATP]P_{(i)}+k_rP_{(ii)}+k_{dMT}P_{(iv)}+k_aP_{(v)}\nonumber\\
\frac{dP_{(ii)}}{dt}&=-(k_r+k_D)P_{(ii)}+k_{bi}[ATP]P_{(i)}\nonumber\\
\frac{dP_{(ii')}}{dt}&=-k_{dADP}P_{(ii')}+k_DP_{(ii)}\nonumber\\
\frac{dP_{(iii)}}{dt}&=-(k_h+k_{bi}^{(iii)}[ATP])P_{(iii)}+k_{dADP}P_{(ii')}+k_r^{(iii)}P_{(iii')}\nonumber\\
\frac{dP_{(iv)}}{dt}&=-(k_{dMT}+k_{bi}^{(iv)}[ATP])P_{(iv)}+k_hP_{(iii)}+k_r^{(iv)}P_{(iv')}\nonumber\\
\frac{dP_{(iii')}}{dt}&=-(k_r^{(iii)}+k^{(iii)}_{diss})P_{(iii')}+k_{bi}^{(iii)}[ATP]P_{(iii)}\nonumber\\
\frac{dP_{(iv')}}{dt}&=-(k_r^{(iv)}+k^{(iv)}_{diss})P_{(iv')}+k_{bi}^{(iv)}[ATP]P_{(iv)}\nonumber\\
\frac{dP_{(v)}}{dt}&=-k_aP_{(v)}+k_{diss}^{(iii)}P_{(iii)}+k_{diss}^{(iv)}P_{(iv)},
\tag{10}
\end{align}
where $P_{(\alpha)}$ is the probability of finding the molecule in a mechanochemical state $\alpha(=i, ii, \cdots v)$ with $\sum_{\alpha} P_{(\alpha)}=1$. 
The steady state solutions by setting $\frac{dP_{(\alpha)}}{dt}=0$ leads to    
\begin{align}
     P_{(i)}&=\frac{1}{k_{bi}[ATP]}\left(1+\frac{k_r}{k_D}\right)\frac{\mathcal{X}}{\mathcal{Z}},\quad P_{(ii)}=\frac{1}{k_D}\frac{\mathcal{X}}{\mathcal{Z}},\quad P_{(ii')}=\frac{1}{k_{dADP}}\frac{\mathcal{X}}{\mathcal{Z}}\nonumber\\
     P_{(iii)}&=\frac{\mathcal{Y}}{\mathcal{Z}},\quad\quad P_{(iv)}=\frac{1}{\mathcal{Z}},\quad\quad P_{(iii')}=K_{m}^{(iii)}\frac{\mathcal{Y}}{\mathcal{Z}},\quad\quad P_{(iv')}=K_{m}^{(iv)}\frac{1}{\mathcal{Z}}\nonumber\\
     P_{(v)}&=\frac{1}{k_a}\left(k_{diss}^{(iii)}K_{m}^{(iii)}\mathcal{Y}+k_{diss}^{(iv)}K_{m}^{(iv)}\right)\frac{1}{\mathcal{Z}}\nonumber\\
     \mathcal{X}&\equiv k_{dMT}\left(1+\frac{k^{(iii)}_{diss}}{k_h}K_{m}^{(iii)}\right)\left(1+\frac{k^{(iv)}_{diss}}{k_{dMT}}K_{m}^{(iv)}\right)\nonumber\\ 
     \mathcal{Y}&\equiv \frac{k_{dMT}}{k_h}\left(1+\frac{k_{diss}^{(iv)}}{k_{dMT}}K_{m}^{(iv)}\right)\nonumber\\ 
     \mathcal{Z}&\equiv\left[1+\left(1+\frac{k_{diss}^{(iii)}}{k_a}\right)K_{m}^{(iii)}\right]\mathcal{Y}+\left[1+\left(1+\frac{k_{diss}^{(iv)}}{k_a}\right)K_{m}^{(iv)}\right]\nonumber\\
&+\left[\frac{1}{k_{bi}[ATP]}\left(1+\frac{k_r}{k_D}\right)+k_D^{-1}+k_{dADP}^{-1}\right]\mathcal{X}\nonumber\\
     K_{m}^{(iii)}&\equiv\frac{k_{bi}^{(iii)}[ATP]}{k_r^{(iii)}+k_{diss}^{(iii)}},\quad K_{m}^{(iv)}\equiv\frac{k_{bi}^{(iv)}[ATP]}{k_r^{(iv)}+k_{diss}^{(iv)}}.
\label{eqn:SS}
\tag{11}
\end{align}
When the average velocity at steady state is computed using $v=d(k_{bi}[ATP]P_{(i)}-k_r P_{(ii)})=\mathcal{X}/\mathcal{Z}$, one can write the velocity in the form of Michaelis-Menten equation.  
\begin{equation}
v=d\frac{\frac{k^*}{(1+\mathcal{Q}([ATP]))}[ATP]}{\frac{\frac{k^*}{1+\mathcal{Q}([ATP])}(1+\frac{k_r}{k_D})}{k_{bi}}+[ATP]}
=d\frac{u_1^o[ATP]}{\frac{u_1^0+w_1^0}{k_0^0}+[ATP]}
=\frac{V_{max}[ATP]}{K_M+[ATP]}
\label{eqn:velocity}
\tag{12}
\end{equation}
where $(k^*)^{-1}=k_D^{-1}+k_{dADP}^{-1}+k^{-1}_{dMT}+k^{-1}_h$, 
\begin{align}
\mathcal{Q}([ATP])&\equiv\frac{k^*}{k_h}\left[\frac{1+\left(1+\frac{k_{diss}^{(iii)}}{k_a}\right)K_m^{(iii)}}{1+\frac{k_{diss}^{(iii)}}{k_h}K_m^{(iii)}}-1\right]\nonumber\\
&+\frac{k^*}{k_{dMT}}\left[\frac{1+\left(1+\frac{k_{diss}^{(iv)}}{k_a}\right)K_m^{(iv)}}{\left(1+\frac{k_{diss}^{(iii)}}{k_h}K_m^{(iii)}\right)\left(1+\frac{k_{diss}^{(iv)}}{k_h}K_m^{(iv)}\right)}-1\right],\nonumber
\tag{13}
\end{align}
and $d=8.2$ $nm$ (the gap between neighboring tubulin binding sites). 
If the dissociation from the microtubule is suppressed by small $K_m^{(iii)}$ and $K_m^{(iv)}$ then 
$\mathcal{Q}\rightarrow 0$.  
Depending on the rate constant, $\mathcal{Q}$ can be either positive or negative. 
Although the large dissociation constant reduces the processivity,  
the presence of dissociation can increase the effective velocity of kinesin if $\mathcal{Q}<0$. 
The second and the third expressions following the equality sign are given to compare our result with the (N=2)-model of Fisher et. al. \cite{Fisher01PNAS} and Michaelis-Menten kinetics, respectively. If $\mathcal{Q}=0$ the (N=2)-model analysis on the experimental data by Block and coworkers \cite{Visscher99Science} predicts  
$k^*=u_1^o=108 s^{-1}$, $k_{bi}=k_0^0=1.80\mu M^{-1}s^{-1}$, and 
$k^*k_r/k_D=w_1^0=6.0s^{-1}$ ($k_D=18\times k_r$). 
This sets the lower bound for the parameters as $k_D$, $k_{dADP}$, $k_{dMT}$, $k_h > 108$ $s^{-1}$. 

The average run length of kinesin, $L$, is calculated using 
\begin{equation}
L=d\times\langle l\rangle=d\times \sum_{l=1}^{\infty}l(1-P_{(v)})^lP_{(v)}=d\times\frac{1-P_{(v)}}{P_{(v)}},
\tag{14}
\end{equation} 
where $l$ is the number of mechanical steps of kinesin. 
If the probability of the dissociated kinesin is small ($P_{(v)}\approx 0$), (i.e., 
if $[(ADP)_Y-(ADP)_X]_{(v)}$ is negligible in Fig. 1), 
then $L\approx d/P_{(v)}=v/k_{diss}$ where $k_{diss}$ is the dissociation rate. 
\\

\clearpage

\section{SI Figure Captions}

{{\bf Fig. 6}: }
Kinesin structure. 
({\it A}) Top view when kinesin is bound to tubulin. The helices on the top with respect to $\beta$-sheet are colored in pink. 
({\it B}) Bottom view. The helices on the bottom with respect to $\beta$-sheet or on the side of the tubulin binding interface are colored in lightblue. 
L11 loop, which is not observable in the crystal structure because of disorder, is tentatively drawn with a dashed line. 
({\it C}) Side view. Note that the neck-linker ($\beta9$, $\beta 10$) is connecting the 
$\alpha 7$ neck-helix with the $\alpha 6$ helix in motor domain. 
({\it D}) View around the nucleotide binding site. 
P-loop, switch-1, switch-2, and N4 regions, which are relevant to nucleotide binding, are colored in red with annotation. \\

{{\bf Fig. 7} :}
Analysis of the kinesin equilibrium dynamics using the Gaussian network model (GNM). 
({\it A}) The cross-correlation map of the residue fluctuation for the two-headed kinesin structure is shown in Fig.2C.  
({\it B}) The amplitudes are color-coded based on its value. 
The relative difference between the trailing and the leading head with respect to leading head ($\delta_{ij}$) is plotted on the right. 
({\it C}) Mean square displacement of residues in the trailing head (red) and the leading head (blue) are plotted on the same plot. 
A comparison of the amplitudes shows that the leading head fluctuates more than the trailing head. 
The relative difference between the two plots with respect to leading head ($\delta_{ii}$) is plotted on the right panel. \\

{{\bf Fig. 8} :}
Analysis of the kinesin equilibrium dynamics using an equilibrium ensemble generated from the simulations under SB-Hamiltonian. 
The legends for {\it A}-{\it C} are identical to SI Fig. 7. \\

{{\bf Fig. 9} :}
Results of the strain induced regulation in the kinesin dimer on the MT generated using the SOP potential (Eq.\ref{eq:SOP}). 
Comparisons between the results obtained with the SOP potential and with the SB potential confirm qualitatively the identical conclusions. \\
\clearpage

\begin{figure}[ht]
\includegraphics[width=6.50in]{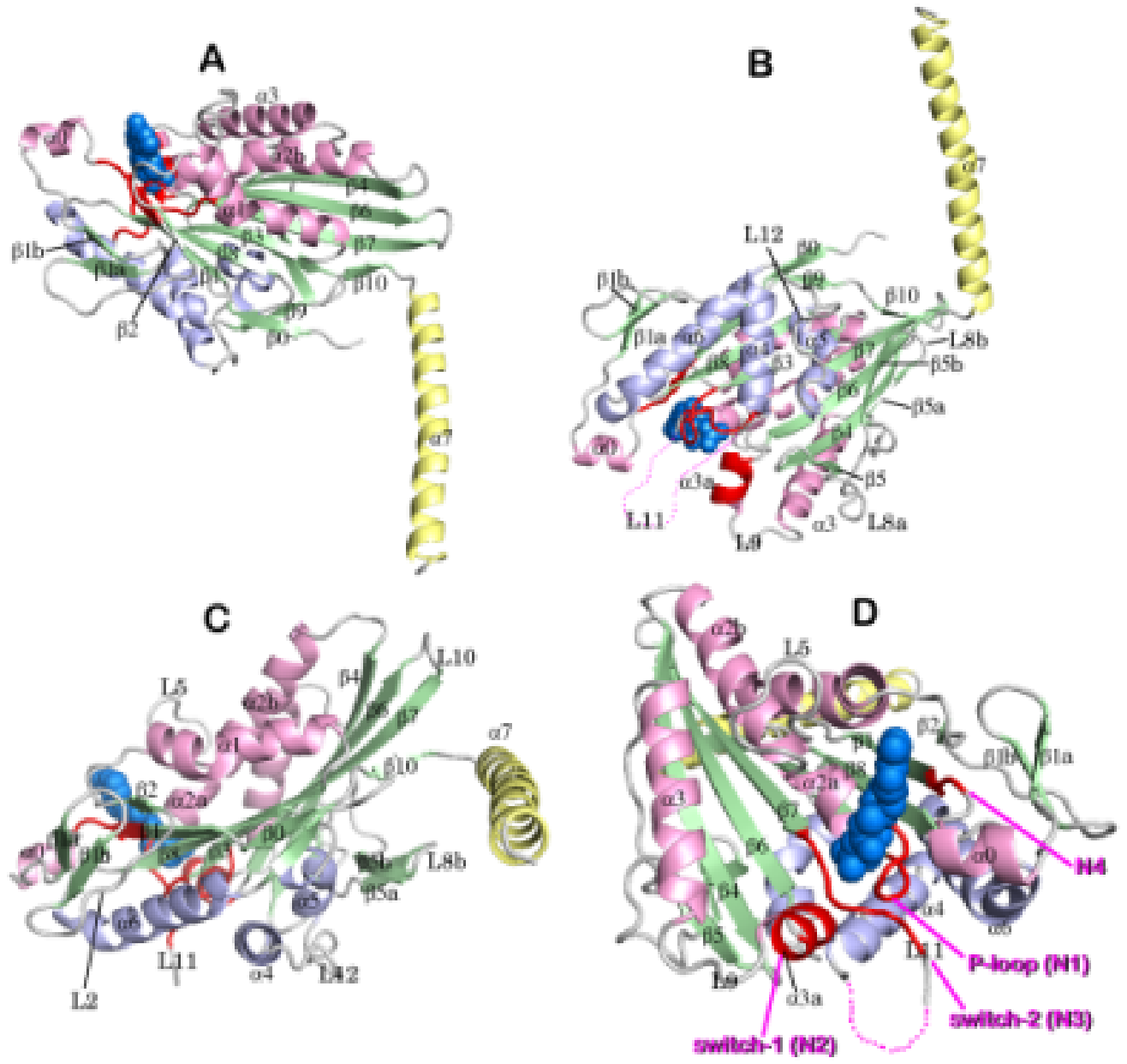}
\caption{}
\end{figure}
\clearpage

\begin{figure}[ht]
\includegraphics[width=7.00in]{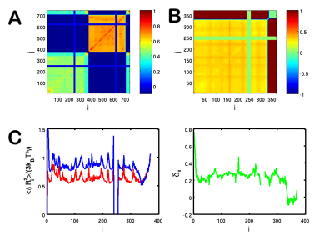}
\caption{}
\end{figure}
\clearpage
\begin{figure}[ht]
\includegraphics[width=7.00in]{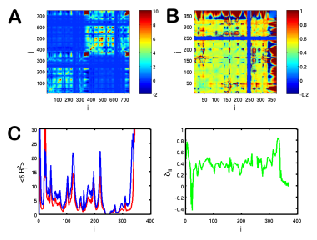}
\caption{}
\end{figure}
\clearpage

\begin{figure}[ht]
\includegraphics[width=7.00in]{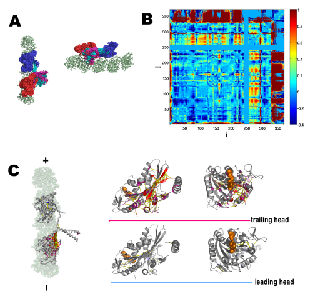}
\caption{}
\end{figure}
\clearpage
\end{document}